\documentclass[twocolumn,aps,prd,superscriptaddress,nofootinbib,floatfix]{revtex4}

\usepackage{graphicx,multirow}
\usepackage{hyperref}

\newcommand{\nn}{\nonumber}
\newcommand{\beq}{\begin{equation}}
\newcommand{\eeq}{\end{equation}}
\newcommand{\beqa}{\begin{eqnarray}}
\newcommand{\eeqa}{\end{eqnarray}}

\def\d{{\rm d}}
\newcommand{\Gammaa}{\Gamma}

\newcommand{\Bbar}{\,\overline{\!B}{}}
\newcommand{\Dbar}{\,\overline{\!D}{}}
\newcommand{\Kbar}{\,\overline{\!K}{}}
\def\B0bar{\Bbar{}^0}
\def\D0bar{\Dbar{}^0}
\def\K0bar{\Kbar{}^0}

\begin{document}

\title{\boldmath Constraining $CP$ violation in neutral meson mixing with theory
input}

\author{Marat Freytsis}
\affiliation{Ernest Orlando Lawrence Berkeley National Laboratory,
University of California, Berkeley, CA 94720}
\affiliation{Berkeley Center for Theoretical Physics, Department of Physics,
University of California, Berkeley, CA 94720}

\author{Zoltan Ligeti}
\affiliation{Ernest Orlando Lawrence Berkeley National Laboratory,
University of California, Berkeley, CA 94720}

\author{Sascha Turczyk}
\affiliation{Ernest Orlando Lawrence Berkeley National Laboratory,
University of California, Berkeley, CA 94720}

\begin{abstract}

There has been a lot of recent interest in the experimental hints of $CP$
violation in $B_{d,s}^0$ mixing, which would be a clear signal of beyond the
standard model physics (with higher significance).  We derive a new relation for
the mixing parameters, which allows clearer interpretation of the data in models
in which new physics enters in $M_{12}$ and/or $\Gamma_{12}$.  Our results
imply that the central value of the D\O\ measurement of the semileptonic $CP$
asymmetry in $B_{d,s}^0$ decay is not only in conflict with the standard model,
but in a stronger tension with data on $\Delta\Gamma_s$ than previously
appreciated.  This result can be used to improve the constraint on
$\Delta\Gamma$ or $A_{\rm SL}$, whichever is less precisely measured.

\end{abstract}

\maketitle

\section{Introduction}

Recently, $CP$ violation in neutral meson mixing received renewed attention due
to the D\O\ hint of $CP$ violation in $B$\,--\,$\Bbar$ mixing, measured
by the $CP$ asymmetry in decays of a $b\bar b$ pair to two same-sign
muons~\cite{Abazov:2011yk},
\beq\label{expD0}
A_{\rm SL}^b = -[7.87 \pm 1.72\, \mbox{(stat)} \pm 0.93\,
\mbox{(syst)}] \times 10^{-3} \,.
\eeq
At the Tevatron both $B^0_d$ and $B^0_s$ are produced, and hence $A_{\rm SL}^b$
is a linear combination of the two asymmetries~\cite{Abazov:2011yk}
\beq\label{weight}
A_{\rm SL}^b = (0.594\pm0.022)\, A_{\rm SL}^d
  + (0.406\pm0.022)\, A_{\rm SL}^s \,.
\eeq
The central value in Eq.~(\ref{expD0}) would be a clear sign of new physics
(NP)~\cite{Laplace:2002ik, Ligeti:2010ia}.  Measurements at the $e^+e^-$ $B$
factories~\cite{Asner:2010qj} and at D\O~\cite{Abazov:2009wg} yield
\beq\label{ASLds}
A_{\rm SL}^d = - (0.5 \pm 5.6) \times 10^{-3}\,,\qquad
A_{\rm SL}^s = - (1.7 \pm 9.2) \times 10^{-3}\,.
\eeq

In $B_s^0$ mixing, a nonzero lifetime difference, $\Delta\Gamma_s \equiv
\Gamma_L-\Gamma_H$, was established recently,
\beqa\label{deltaGs}
\Delta\Gamma_s &=& (0.116 \pm 0.019)\, {\rm ps}^{-1}\,, \qquad
  \mbox{LHCb~\cite{lhcb}}, \nn\\
\Delta\Gamma_s &=& (0.068 \pm 0.027)\, {\rm ps}^{-1}\,, \qquad
  \mbox{CDF~\cite{CDF:2011af}}, \nn\\
\Delta\Gamma_s &=& (0.163^{+0.065}_{-0.064})\, {\rm ps}^{-1}\,, \qquad\quad
  \mbox{D\O~\cite{Abazov:2011ry}}.
\eeqa
In the absence of a world average, we use the most precise measurement from
LHCb.  For $\Delta m_s$ the average of the CDF~\cite{Abulencia:2006ze} and
LHCb~\cite{LHCbmixing} measurements is
\beq\label{deltams}
\Delta m_s\equiv m_H-m_L = (17.731 \pm 0.045)\, {\rm ps}^{-1} .
\eeq

One should naturally ask if there are any constraints on the mixing parameters,
beyond the obvious one: that the mass and width eigenvalues of the heavy and
light mass eigenstates, $m_{H,L}$ and $\Gamma_{H,L}$, must be positive.  (We use
the notation customary in $B$ physics, but the results apply equally for $K^0$
and $D^0$ mixing as well.) The time evolution of the flavor eigenstates is
\beq\label{timedep}
i\, {\d\over \d t} 
  \pmatrix{|B^0(t)\rangle\cr |\B0bar(t)\rangle} 
= \bigg(M - {i\over2}\,\Gamma\bigg)
  \pmatrix{|B^0(t)\rangle\cr |\B0bar(t)\rangle} ,
\eeq
where $M$ and $\Gamma$ are $2\times2$ Hermitian matrices, and $CPT$ invariance
implies $M_{11} = M_{22}$ and $\Gamma_{11} = \Gamma_{22}$.
The physical states are the eigenvectors of the Hamiltonian,
\beq\label{physeigen}
|B_{H,L}\rangle = p\, |B^0\rangle \mp q\, |\B0bar\rangle \,,
\eeq
where we chose $|p|^2+|q|^2=1$.  $CP$ violation in mixing occurs if the mass and
$CP$ eigenstates do not coincide,
\beq\label{cpmix}
\delta \equiv \langle B_H | B_L \rangle = \frac{|p|^2-|q|^2}{|p|^2+|q|^2} 
  = {1-|q/p|^2\over 1+|q/p|^2} \neq 0\,.
\eeq
The solution for the mixing parameters satisfies
\beq\label{qpratio}
\frac{q^2}{p^2} = \frac{2M_{12}^* - i\Gamma_{12}^*}{2M_{12} - i\Gamma_{12}}\,,
\eeq
and from this and Eq.~(\ref{cpmix}) it follows that (see, e.g.,
\cite{Branco:1999fs})
\beq\label{simplebound}
\delta < {\rm min} \bigg( \frac{|2M_{12}|}{|\Gamma_{12}|}\,, \
  \frac{|\Gamma_{12}|}{|2M_{12}|} \bigg)\,.
\eeq
The measurable CP asymmetry in semileptonic (or any ``flavor-specific") decay
can be expressed as
\beq\label{cpasl}
A_{\rm SL} = {1-|q/p|^4\over 1+|q/p|^4} = \frac{2 \delta}{1+\delta^2}
  = \frac{{\rm Im}\, (\Gamma_{12}/ M_{12})}
  {1 + |\Gamma_{12}|^2/(4\, |M_{12}|^2)}\,.
\eeq
Thus, in the small $\delta$ limit, $A_{\rm SL} = 2 \delta + {\cal O}(\delta^3)$.

In the $|\Gamma_{12}/ M_{12}| \ll 1$ limit, which applies model independently
for the $B_{d,s}^0$ systems,
\beqa\label{approx}
\Delta m &=& 2\, |M_{12}|\,
  \big[ 1 + {\cal O}(|\Gamma_{12}/ M_{12}|^2) \big], \nn\\
\Delta\Gamma &=& 2\, |\Gamma_{12}|\, \cos [{\rm arg}(-\Gamma_{12}/M_{12})]\, 
  \big[ 1 + {\cal O}(|\Gamma_{12}/ M_{12}|^2) \big], \nn\\
A_{\rm SL} &=& \textrm{Im}\,(\Gamma_{12}/M_{12})\,
  \big[ 1 + {\cal O}(|\Gamma_{12}/ M_{12}|^2) \big].
\eeqa
In this limit, Eq.~(\ref{qpratio}) implies that $q/p$ is a pure phase to a good
approximation, determined by $M_{12}$, which has good sensitivity to NP. 
However, if $|\Gamma_{12} / M_{12}| = {\cal O}(1)$, relevant for $K^0$ and
$D^0$ mesons, then $q/p$ depends on both $\Gamma_{12}$ and $M_{12}$ and the
sensitivity to NP in $M_{12}$ (and in ${\rm arg}\, M_{12}$) is
diluted~\cite{Bergmann:2000id}. In that case Eq.~(\ref{approx}) does not hold,
but $A_{\rm SL} = 2 \delta$ is a good approximation even in the $D^0$ system,
where the current bound is $|\delta| \lesssim 0.2$~\cite{Asner:2010qj}.

An additional constraint, the unitarity bound~\cite{Bell:1990tq,Lee:1965hi},
stems from the time-evolution of the normalization of any linear combination of
$|B^0\rangle$ and $|\Bbar^0\rangle$ being determined entirely by the $\Gamma$
matrix. As discussed below, this constrains the eigenvalues of $\Gamma$ to be
positive definite independent of the physical eigenvalues, or equivalently
\beq\label{unitbound}
\delta^2 < \frac{\Gamma_H \Gamma_L} {(m_H-m_L)^2 + (\Gamma_H + \Gamma_L)^2/4}
  = \frac{1-y^2}{1+x^2}\,.
\eeq
Here we define, using $\Gammaa = (\Gamma_H+\Gamma_L)/2$,
\beq
  x = \frac{m_H - m_L}{\Gammaa}\,, \qquad
  y = \frac{\Gamma_L - \Gamma_H}{2\Gammaa}\,,
\eeq
where $x$ is positive by definition, while $y \in (-1,\, +1)$.

One may ask if other constraints exist purely from consistency considerations.
As we show in a separate paper~\cite{TBA}, no additional limit on $\delta$
exists, either physical or mathematical, without some knowledge of the
Hamiltonian.  In particular, the invalidity of the bounds claimed
in~\cite{Berger:2007ax, Berger:2010wt} can be made apparent by introducing new
short-distance physics, which reduces $\big|\cos[{\rm arg}
(\Gamma_{12}/M_{12})]\big|$ while it leaves $|M_{12}|$ and $\Gamma_{12}$
unchanged.

However, this does not preclude the presence of new relations from appropriate
inclusions of theoretical predictions from models of the underlying
interactions.  In this paper, we derive a relation between the mixing
parameters of a meson system and only the magnitude $|\Gamma_{12}|$, which is
in tension with the D\O\ measurement in Eq.~(\ref{expD0}).  

It has been known that the data in Eqs.~(\ref{expD0}) -- (\ref{ASLds}) is not
only in tension with the SM, but --- assuming the SM calculation of
$\Gamma_{12}$ --- also with all models in which NP enters only through
$M_{12}$~[\onlinecite{Dobrescu:2010rh}, \onlinecite{Ligeti:2010ia}].  This is
because Eqs.~(\ref{simplebound}) and (\ref{approx}) imply $\delta <
|\Gamma_{12}| / \Delta m$. Our result goes beyond this, because it makes optimal
use of data on $\Delta\Gamma$ without theoretical assumptions, and indicates a
larger tension independent of the nature of typical new physics.

\section{Unitarity with theory input}

As mentioned above, Eq.~(\ref{unitbound}) was first derived in 
Refs.~\cite{Bell:1990tq, Lee:1965hi}. We show that a stronger bound on $\delta$
can be obtained using additional input from theoretical calculations. An analogy
to the derivation of Ref.~\cite{Bell:1990tq} will be particularly useful in
deriving our results. 

We define the complex quantities
\beq\label{uboundvec}
a_i = \sqrt{2\pi \rho_i}\, \langle f_i | \mathcal{H} | B \rangle \,, \qquad
\bar{a}_i = \sqrt{2\pi \rho_i}\, \langle f_i | \mathcal{H} | \Bbar \rangle \,,
\eeq
with $\rho_i$ denoting the phase space density for final state $f_i$. If we
treat $a_i$ and $\bar{a}_i$ as vectors in a complex $N$-dimensional vector
space, then taking the standard inner product on complex vector spaces, and
using the optical theorem~\cite{Bell:1990tq}, amounts to the relations
\beq\label{ubound12}
  a_i^*\, a_i = \Gamma_{11}\,, \qquad
  \bar{a}_i^*\, \bar{a}_i = \Gamma_{22} \,, \qquad
  \bar{a}_i^*\, a_i = \Gamma_{12} \,,
\eeq
where $CPT$ fixes $\Gamma_{11} = \Gamma_{22} = \Gamma$. Applying the
Cauchy-Schwarz inequality to the vectors $a_i$ and $\bar{a}_i$
implies~\cite{Bell:1990tq}
\beq\label{gammapos}
  |\Gamma_{12}| \leq \Gamma_{11}\,.
\eeq
This is equivalent to the statement that the eigenvalues of the
$\Gamma$ matrix must be positive (in addition to $\Gamma_{H,L}>0$).

To see that this is also equivalent to the unitarity bound of
Eq.~(\ref{unitbound}), we use Eq.~(\ref{physeigen}) to define new vectors
$a_{Hi}$ and $a_{Li}$ analogously, such that
\beq\label{uboundphysvec}
a_i = \frac{1}{2p}\, (a_{Hi} + a_{Li}) \,,\qquad
\bar{a}_i = \frac{1}{2q}\, (a_{Li} - a_{Hi}) \,.
\eeq
For these newly defined vectors we can derive, in a similar manner as for
Eq.~(\ref{ubound12}), the relations
\beqa\label{uboundHL}
  a_{Hi}^*\, a_{Hi} &=& \Gamma_{H} \,, \qquad
  a_{Li}^*\, a_{Li} = \Gamma_{L} \,, \\
  a_{Hi}^*\, a_{Li} &=& -i (m_H - m_L + i\Gammaa) \, \delta \,. \nn
\eeqa
Substituting Eq.~(\ref{uboundphysvec}) into Eq.~(\ref{ubound12}), using
Eq.~(\ref{uboundHL}) and the $|q/p|^2 = (1-\delta)/(1+\delta)$ identity, we
obtain $\Gamma_{11}$ and $\Gamma_{12}$ in terms of $M_{H,L},\ \Gamma_{H,L}$ and
$\delta$. The unitarity bound in Eq.~(\ref{unitbound}) then arises from using
the new expressions for $\Gamma_{11}$ and $\Gamma_{12}$ in Eq.~(\ref{gammapos}).

The preceding derivation made no assumptions on the actual values of the matrix
elements appearing in the problem. For the kaon system, for which this bound was
originally derived, this was a necessity due to the dominance of long-distance
physics in the result. For $B_{d,s}$ mesons, the large mass scale $m_b \gg
\Lambda_{\rm QCD}$ allows $\Gamma_{11}$ and $\Gamma_{12}$ to be calculated in an
operator product expansion, and at leading order $|\Gamma_{12} / \Gamma_{11}| =
{\cal O}[(\Lambda_{\rm QCD}/m_b)^3\, (16\pi^2)]$, where $16\pi^2$ occurs due to
a one-loop difference between the two calculations.  (In the $B_d$ system there
is an additional CKM suppression.)  Thus it makes sense to consider some theory
input, and we define 
\beq\label{epsdef}
y_{12} = |\Gamma_{12}| \,\big/\, \Gammaa \,.
\eeq 
Using this relation between these matrix elements and proceeding with the same
steps as above, we obtain, instead of an inequality as in Eq.~(\ref{gammapos}),
\beq\label{unitrel}
\delta^2 = \frac{y_{12}^2-y^2}{y_{12}^2+x^2} =
  \frac{|\Gamma_{12}|^2-(\Delta\Gamma)^2/4}{|\Gamma_{12}|^2+(\Delta m)^2}\,. 
\eeq
This equation follows from the solution of the eigenvalue problem, and was
previously derived in Ref.~\cite{Branco:1999fs} with the resulting bound on
$\delta$ noted.\footnote{We were unaware of this in v1 of this paper, and we
thank Luis Lavoura and Joao Silva for bringing this to our attention.} (It also
follows from Eqs.~(9) and (12) in~\cite{Grossman:2009mn}.)  For fixed $x$ and
$y$, $\delta^2$ is monotonic in $y_{12}$, so an upper bound on $y_{12}$ gives an
upper bound on $|\delta|$. For $y_{12} \leq 1$ the usual unitarity bound in
Eq.~(\ref{unitbound}) is recovered.

Equation~(\ref{unitrel}) can also be obtained from a scaling argument: As
$\delta$ only depends on mixing parameters, it is independent of the value of
$\Gamma$. One can then scale $\Gamma$ by $y_{12}$, which cannot affect $\delta$
but changes $x\to x/y_{12}$ and $y \to y/y_{12}$. Eq.~(\ref{unitrel}) follows
then from this argument and Eq.~(\ref{unitbound}). The derivations above make
the physical origin of this relation clear.  Even if $CPT$ is violated, the
scaling argument, and therefore Eq.~(\ref{unitrel}) holds, although
Eq.~(\ref{uboundphysvec}) is modified.  This applies for $|\delta|^2$, as $CPT$
violation allows $\delta$ to be complex.

Even if a precise calculation of $\Gamma_{12}$ is not possible or one assigns a
very conservative uncertainty to it, an upper bound on $y_{12}$ implies an upper
bound on $|\delta|$, which is stronger than that in Eq.~(\ref{unitbound}). For
small values of $y_{12}$, as in the $B_d$ system, this bound can be much
stronger.

\section{Comparing Data and Theory}

We can compare the absolute value of $A_{\rm SL}^b$ measured by D\O, with the
result implied by the relation above. At present Eq.~(\ref{unitrel}) only
provides an upper bound on $|\delta|$, as the uncertainties of $\Gamma_{12}$ and
$\Delta\Gamma$ allow the numerators for both $B_d$ and $B_s$ to vanish. Denoting
this upper bound by $\delta^{d,s}_{\text{max}}$ and using the weight factors 
from Eq.~(\ref{weight}),
\begin{equation}
|A_{\rm SL}^b| \le (1.188\pm 0.044)\, \delta^d_{\text{max}}
  + (0.812\pm0.044)\, \delta^s_{\text{max}} .
\end{equation}
Since Eq.~(\ref{unitrel}) only bounds $|\delta|$, the bound on $A_{\rm SL}^b$ is
not sensitive to possible cancellations between $A_{\rm SL}^d $ and $A_{\rm
SL}^s$ (cf., the opposite signs of $A_{\rm SL}^{s,d}$ in the SM, although
$|A_{\rm SL}^s| \ll |A_{\rm SL}^d|$). As $\Delta m_{d,s}$ are precisely known,
we plot the bound as a function of $\Delta\Gamma_{d,s}$. If LHCb measures
$A_{\text{SL}}^s - A_{\text{SL}}^d$~\cite{ASLdiff}, then the above bound with
modified coefficients apply for that measurement.

\begin{figure}[t]
\includegraphics[width=.9\columnwidth]{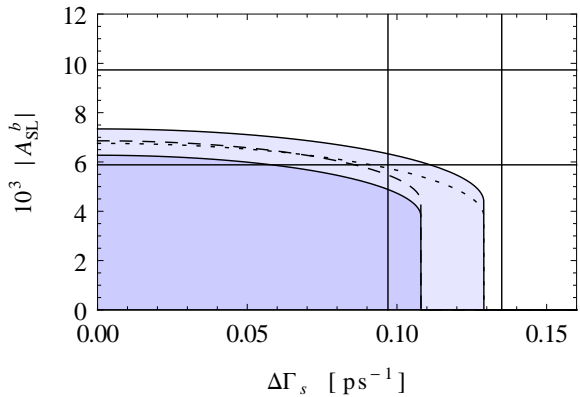}
\caption{Upper bounds on $A_{\rm SL}^b$ as a function of $\Delta\Gamma_s$,
setting $\Delta\Gamma_d = 0$.  The darker [lighter] shaded region is allowed
using the $1\sigma$ [$2\sigma$] range of the theory calculation of
$|\Gamma_{12}^{d,s}|$. The pair of horizontal [vertical] lines show the
$1\sigma$ range of the measured $|A_{\rm SL}^b|$ from D\O\ [$\Delta \Gamma_s$
from LHCb].  The other curves are described in the text.\label{1Dplot}}
\end{figure}

In Fig.~\ref{1Dplot} we set $\Delta \Gamma_d = 0$, which gives the most
conservative bound.  The darker shaded region shows the upper bound on $A_{\rm
SL}^b$ using the $1\sigma$ ranges for $|\Gamma_{12}^{d,s}|$ in the
SM~\cite{Lenz:2011ti}, $2|\Gamma_{12}^s| = (0.087 \pm 0.021)\, {\rm ps}^{-1}$
and $2|\Gamma_{12}^d| = (2.74 \pm 0.51)\times 10^{-3}\, {\rm ps}^{-1}$.  The
dashed [dotted] curve shows the impact of using the $2\sigma$ region for
$\Gamma_{12}^d$ [$\Gamma_{12}^s$], and the lighter shaded region includes both
$2\sigma$ regions.  The vertical boundaries of the shaded regions arise because
$|\Delta\Gamma_s| > 2\, |\Gamma_{12}^s|$ is unphysical. A tension between
the $A_{\rm SL}^b$ measurement and the bound is visible, independent of
the discrepancy between the D\O\ result and the global fit to the
latest available experimental data~\cite{Lenz:2012az}.

\begin{figure*}[tb]
  \includegraphics[width=.9\columnwidth]{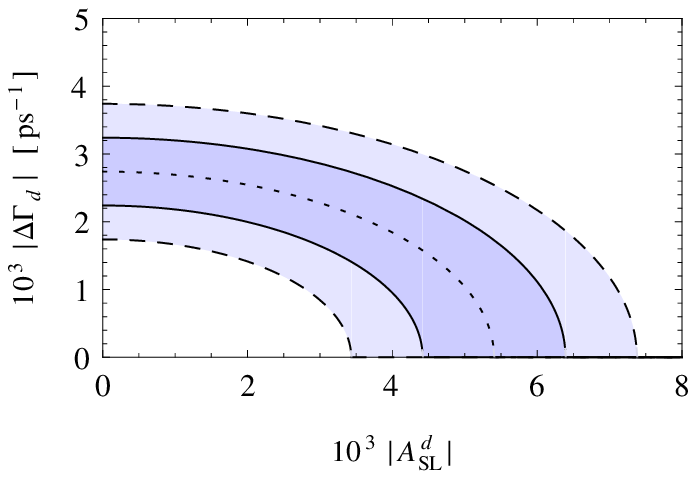} \hfil\hfil
  \includegraphics[width=.9\columnwidth]{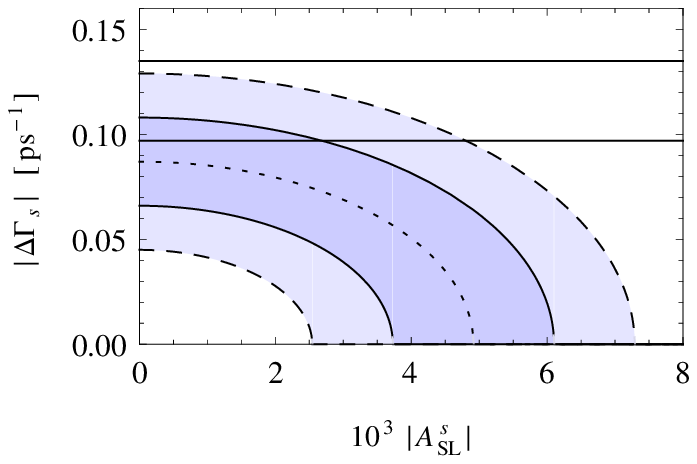}
\caption{Left plot: the region allowed by Eq.~(\ref{unitrel}) in the $A_{\rm
SL}^d - \Delta\Gamma_d$ plane.  The SM calculation of $|\Gamma_{12}^{d,s}|$ at
$1\sigma$ [$2\sigma$] gives the darker [lighter] shaded region.
Right plot: same for $A_{\rm SL}^s - \Delta\Gamma_s$; the straight lines show
the $1\sigma$ range of the LHCb result for~$\Delta\Gamma_s$.\label{2Dplot}}
\end{figure*}

We derived not an absolute bound in the fashion of the unitarity bound but a
relation between calculable and measured quantities.  It is thus worth
clarifying the relationship of our result to the stated $3.9\,\sigma$
disagreement of $A_{\rm SL}^b$ with the SM reported in~\cite{Abazov:2011yk}. 

The SM prediction of $A_{\rm SL}$ uses the calculation of $\Gamma_{12}$, and
$|\Gamma_{12}|$ also enters our bound; thus, the discrepancies are correlated. 
Although the calculation of $|\Gamma_{12}|$ and ${\rm Im}(\Gamma_{12})$ both
rely on the same operator product expansion and perturbation theory, the
existence of large cancellations in ${\rm Im}(\Gamma_{12})$ may lead one to
think that the uncertainties could be larger in its SM calculation than what is
tractable in the behavior of its next-to-leading order
calculation~\cite{Ciuchini:2003ww, Beneke:2003az}. The sensitivity of
$\Gamma_{12}$ to new physics is generally weaker than that of $M_{12}$
(see~\cite{Bai:2010kf, Bobeth:2011st} for other options).  Thus, it is
interesting to determine $\delta$ from Eq.~(\ref{unitrel}), besides its direct
calculation.

Of course, independent of the D\O\ measurement of $A_{\rm SL}^b$, we can also
compare the bound implied by our relation to the individual best bounds on the
semileptonic asymmetries in Eq.~(\ref{ASLds}). To this end, in Fig.~\ref{2Dplot}
we plot $A_{\rm SL}^d$ vs.\ $\Delta\Gamma_d$ (and similarly for $B_s$) allowed
by Eq.~(\ref{unitrel}) and the $1\sigma$ and $2\sigma$ ranges of the SM
calculation of $|\Gamma_{12}|$~\cite{Lenz:2011ti}. Here, there have been no
discrepancies claimed between the theory predictions and measurements, but our
relation allows us to place a bound tighter than the current experimental
constraints which is more robust than the purely theoretical SM calculation as
outlined above.

Using $\Delta\Gamma_s$ from LHCb in Eq.~(\ref{deltaGs}), and neglecting
$\Delta\Gamma_d$, we find the $2\sigma$ level bounds,
\beq\label{newbound}
  |A_{\text{SL}}^d| < 7.4 \times 10^{-3}, \qquad 
  |A_{\text{SL}}^s| < 4.2 \times 10^{-3}. 
\eeq
While this bound on $A_{\text{SL}}^s$ may seem to disagree with
Fig.~\ref{2Dplot}, note that in the plot the uncertainties of $\Gamma^s_{12}$
and $\Delta\Gamma_s$ are not combined.   Propagating the uncertainties,
$|\Gamma_{12}^s|^2-(\Delta\Gamma_s)^2/4$ is negative at the $1\sigma$ level, an
unphysical result, hence the $2\sigma$ bounds in Eq.~(\ref{newbound}).  This
bound on $A_{\text{SL}}^s$ is better than the current best bound in
Eq.~(\ref{ASLds}) by more than a factor of 3, while that for $A_{\text{SL}}^d$
is comparable.  (However, in the case of $B_d$ this is driven primarily by the
uncertainty in the lifetime difference. If a non-zero value of $\Delta\Gamma_d$
were observed, a better bound could be derived.) It is worth emphasizing that
this implication goes in both directions, given that an observation of
$A_{\text{SL}}^d \neq 0$ may happen before that of $\Delta\Gamma_d \neq 0$.  Due
to Eq.~(\ref{unitrel}), as soon as one of the two is measured to be nonzero, the
other is constrained to be significantly smaller at worst and given a definite
prediction at best.

\section{conclusions}

We provided a physical derivation of Eq.~(\ref{unitrel}) for neutral meson
oscillation parameters, especially relevant for the $B_{d,s}^0$ systems, which
allows incorporating theoretical input on $|\Gamma_{12}|$ without any
approximation, and with or without $CPT$ conservation. This input is typically
insensitive to the nature of NP and avoids the largest uncertainties of the
direct theoretical calculation of $CP$ violation in mixing. The application to
the two neutral $B$ systems, taking into account the recent LHCb measurement
\cite{lhcb}, leads to bounds on the semileptonic $CP$ asymmetries of both
system. These bounds are in tension with the D\O\ measurement of $A_{\rm SL}^b$,
while providing a bound on the individual asymmetries at comparable or better
levels than the current experimental bounds. Additionally, once an unambiguous
determination of $A_{\rm SL}$ or $\Delta\Gamma$ is made, we can use it to
constrain the other observable.  Refinements of the $A_{\rm SL}^{d,s}$
measurements are an important part of the future $B$ physics
program~\cite{Ligeti:2006pm, Grossman:2009dw} to search for new physics at both
LHCb and the $e^+e^-$ $B$ factories. Future bounds will in particular be helpful
to constrain the individual measurements of $A_{\rm SL}$ against the SM as well
as consistency checks.

\begin{acknowledgments}

We thank Cliff Cheung for not entirely useless discussions, Aneesh Manohar for
raising the issue of $CPT$ conservation, and Yuval Grossman and Yossi
Nir for helpful conversations when Ref.~\cite{Berger:2010wt} appeared.
We thank Doug Tuttle and Lynn Brantley for organizing the first BCTP summit
at Glenbrook, NV, where some of these results were obtained.  (Special thanks
for the golf carts, for inspiration.)  This work was supported in part by
the Director, Office of Science, Office of High Energy Physics of the U.S.\
Department of Energy under contract DE-AC02-05CH11231. ST~is supported by a DFG
Forschungsstipendium under contract no.~TU350/1-1. 

\end{acknowledgments}


\begin{thebibliography}{99}

\bibitem{Abazov:2011yk}
  V.~M.~Abazov {\it et al.} [D\O\ Collaboration],
  Phys.\ Rev.\  {\bf D84}, 052007 (2011).
  [arXiv:1106.6308 [hep-ex]].

\bibitem{Laplace:2002ik}
  S.~Laplace, Z.~Ligeti, Y.~Nir, G.~Perez,
  Phys.\ Rev.\  {\bf D65}, 094040 (2002).
  [hep-ph/0202010].

\bibitem{Ligeti:2010ia}
  Z.~Ligeti, M.~Papucci, G.~Perez, J.~Zupan,
  Phys.\ Rev.\ Lett.\  {\bf 105}, 131601 (2010).
  [arXiv:1006.0432 [hep-ph]].

\bibitem{Asner:2010qj}
  D.~Asner {\it et al.}  [Heavy Flavor Averaging Group],
  arXiv:1010.1589 [hep-ex];
  and updates at \url{http://www.slac.stanford.edu/xorg/hfag/}

\bibitem{Abazov:2009wg} 
  V.~M.~Abazov {\it et al.}  [D0 Collaboration],
  Phys.\ Rev.\ D {\bf 82}, 012003 (2010)
  [Erratum-ibid.\ D {\bf 83}, 119901 (2011)]
  [arXiv:0904.3907 [hep-ex]].

\bibitem{lhcb}
LHCb Collaboration, CERN-LHCb-CONF-2012-002,
\url{https://cdsweb.cern.ch/record/1423592}.

\bibitem{CDF:2011af}
M.\ Dorigo, CDF Collaboration, Talk at Moriond QCD 2012,
\url{http://moriond.in2p3.fr/QCD/2012/qcd.html}.

\bibitem{Abazov:2011ry}
  V.~M.~Abazov {\it et al.}  [D0 Collaboration],
  Phys.\ Rev.\  D {\bf 85} (2012) 032006
  [arXiv:1109.3166 [hep-ex]].

\bibitem{Abulencia:2006ze}
  A.~Abulencia {\it et al.} [ CDF Collaboration ],
  Phys.\ Rev.\ Lett.\  {\bf 97}, 242003 (2006).
  [hep-ex/0609040].

\bibitem{LHCbmixing}
The LHCb Collaboration, LHCb-CONF-2011-050.

\bibitem{Branco:1999fs}
  G.~C.~Branco, L.~Lavoura, J.~P.~Silva,
  Int.\ Ser.\ Monogr.\ Phys.\  {\bf 103}, 1-536 (1999).
  See in particular Ch.~6 and 30.

\bibitem{Bergmann:2000id}
  S.~Bergmann, Y.~Grossman, Z.~Ligeti, Y.~Nir and A.~A.~Petrov,
  Phys.\ Lett.\  B {\bf 486}, 418 (2000)
  [hep-ph/0005181].

\bibitem{Bell:1990tq}
J.S. Bell, J. Steinberger, ``Weak interactions of kaons", in R.~G.~Moorhouse et
al., Eds., Proceedings of the Oxford Int.\ Conf.\ on Elementary Particles,
Rutherford Laboratory, Chilton, England, 1965, p.~195.

\bibitem{Lee:1965hi}
  T.~D.~Lee, L.~Wolfenstein,
  Phys.\ Rev.\  {\bf 138}, B1490-B1496 (1965).

\bibitem{TBA}
  M.~Freytsis, Z.~Ligeti, S.~Turczyk,
  to appear.

\bibitem{Berger:2007ax}
  C.~Berger, L.~Sehgal,
  Phys.\ Rev.\  {\bf D76}, 036003 (2007).
  [arXiv:0704.1232 [hep-ph]].

\bibitem{Berger:2010wt}
  C.~Berger, L.~M.~Sehgal,
  Phys.\ Rev.\  {\bf D83}, 037901 (2011).
  [arXiv:1007.2996 [hep-ph]].

\bibitem{Dobrescu:2010rh} 
  B.~A.~Dobrescu, P.~J.~Fox and A.~Martin,
  Phys.\ Rev.\ Lett.\  {\bf 105}, 041801 (2010)
  [arXiv:1005.4238 [hep-ph]].

\bibitem{Grossman:2009mn} 
  Y.~Grossman, Y.~Nir and G.~Perez,
  Phys.\ Rev.\ Lett.\  {\bf 103}, 071602 (2009)
  [arXiv:0904.0305 [hep-ph]].

\bibitem{ASLdiff}
See, e.g., M.~Calvi, Talk at FPCP~2011 (p.\,29),
\url{http://physics.tau.ac.il/fpcp2011}.

\bibitem{Lenz:2011ti} 
  A.~Lenz and U.~Nierste,
  arXiv:1102.4274 [hep-ph].

\bibitem{Lenz:2012az}
  A.~Lenz {\it et al.},
  arXiv:1203.0238 [hep-ph].

\bibitem{Ciuchini:2003ww} 
  M.~Ciuchini, E.~Franco, V.~Lubicz, F.~Mescia and C.~Tarantino,
  JHEP {\bf 0308}, 031 (2003)
  [hep-ph/0308029].

\bibitem{Beneke:2003az} 
  M.~Beneke, G.~Buchalla, A.~Lenz and U.~Nierste,
  Phys.\ Lett.\ B {\bf 576}, 173 (2003)
  [hep-ph/0307344].

\bibitem{Bai:2010kf} 
  Y.~Bai and A.~E.~Nelson,
  Phys.\ Rev.\ D {\bf 82}, 114027 (2010)
  [arXiv:1007.0596 [hep-ph]].

\bibitem{Bobeth:2011st}
  C.~Bobeth and U.~Haisch,
  arXiv:1109.1826 [hep-ph].

\bibitem{Ligeti:2006pm}
  Z.~Ligeti, M.~Papucci, G.~Perez,
  Phys.\ Rev.\ Lett.\  {\bf 97}, 101801 (2006).
  [hep-ph/0604112].

\bibitem{Grossman:2009dw}
  Y.~Grossman, Z.~Ligeti, Y.~Nir,
  Prog.\ Theor.\ Phys.\  {\bf 122}, 125-143 (2009).
  [arXiv:0904.4262 [hep-ph]].

\end{thebibliography}
\end{document}